# Narrow Band Continuum Colors of Distant Cluster Populations

J. SCHOMBERT[1], K. RAKOS[2]
*(1) Dept. of Physics, University of Oregon*
*(2) Institute for Astronomy, University of Vienna*

## Abstract

In this poster, we present new results on narrow band photometry for A2218 ($z$=0.18) and A2125 ($z$=0.25), two clusters with similar redshifts, but very different cluster properties. A2218 is a dense, elliptical-rich cluster (Bautz-Morgan type II) similar to Coma in its evolutionary appearance. A2125 is a less dense, more open cluster (Bautz-Morgan type II-III), although similar in richness. The color-magnitude relation indicates that A2125 has a more developed blue population than A2218 (the Butcher-Oemler effect), although both clusters have significant numbers of blue galaxies (normal star-forming or starburst) compared to a present-day cluster. The red population are identical in A2125 and A2218 and well fit by passive evolution models.

## 1.1    Introduction

Photometry of distant cluster populations addresses many issues in galaxy evolution studies. Color evolution, Butcher-Oemler effect, cD galaxy construction and S0 galaxy formation are all processes visible through detailed photometry. In recent years, spectroscopy has gained dominance in cluster studies (due to the growth in multi-fiber spectrographs) providing line indices measurements and kinematic information. However, photometry still has its advantages of deeper limiting magnitude and spatial resolution while maintaining spectrophotometric quality with properly selected filter sets.

For the past decade, we have used a set of narrow band filters around the 4000Å break (a modified Strömgren system) 'redshifted' to each cluster's velocity to avoid k-corrections and provide for photometric membership selection. These datasets compliment the work of spectral studies. For example, our narrow band colors focusing on the mean star formation rates averaged over the past Gyr rather than instantaneous rates giving by emission lines. The Strömgren system consists of four filters ($uz,vz,bz,yz$, shown above) which are centered around the 4000Å break. The $bz$ and $yz$ filters ($\lambda_{eff}$ = 4675Å and 5500Å), sample the relatively line free region longward of 4600Å and produce a temperature color index, $bz-yz$. The second region is a band shortward of 4600Å, but above the Balmer discontinuity and is exploited by the $vz$ filter ($\lambda_{eff}$ = 4100Å). This region is strongly influenced by metal absorption lines (i.e. Fe, CN) particularly for spectral classes F to M which dominate the contribution of light in old stellar populations. The third region is a band shortward of the Balmer discontinuity or below the effective limit of crowding of the Balmer absorption lines. This region is explored by the $uz$ filter ($\lambda_{eff}$ = 3500Å).

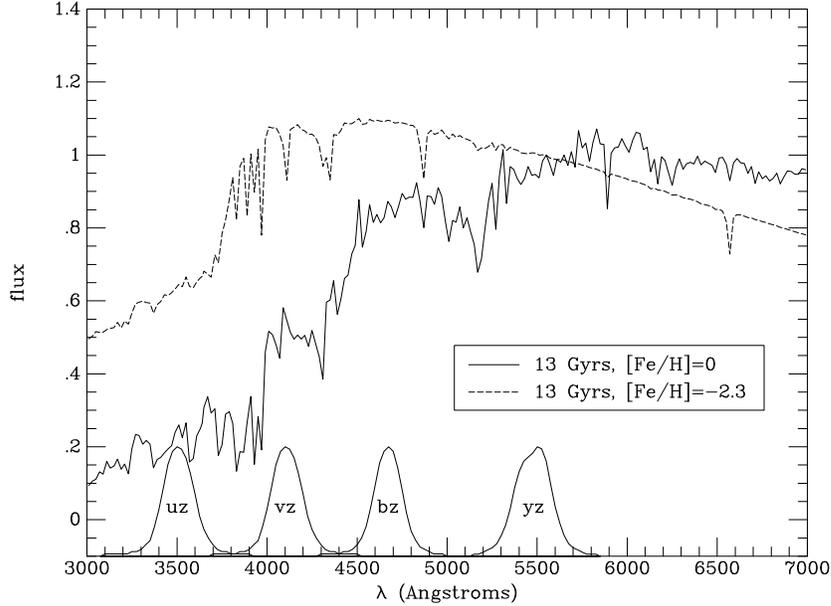

Fig. 1.1. The Strömgren system consists of four filters (*uz,vz,bz,yz*, shown above) which are centered around the 4000Å break. The *bz* and *yz* filters ($\lambda_{eff} = 4675$Å and 5500Å), sample the relatively line free region longward of 4600Å and produce a temperature color index, $bz-yz$. The second region is a band shortward of 4600Å, but above the Balmer discontinuity and is exploited by the *vz* filter ($\lambda_{eff} = 4100$Å). This region is strongly influenced by metal absorption lines (i.e. Fe, CN) particularly for spectral classes F to M which dominate the contribution of light in old stellar populations. The third region is a band shortward of the Balmer discontinuity or below the effective limit of crowding of the Balmer absorption lines. This region is explored by the *uz* filter ($\lambda_{eff} = 3500$Å).

The advantages to the Strömgren system for galaxy evolution are: (1) Cluster membership is assigned based on photometric criteria, thus the ability to examine the lower luminosity galaxies in distant clusters that spectroscopy can not reach. (2) The mean star formation rate is defined by spectrophotometric criteria, as compared to morphological appearance. Principal moment analysis allows us to divided our samples into E (old), S- (passive, S0-type), S (disk star formation rates) and S+ (starburst, high star formation rates). These spectrophotometric classifications are used to study the blue population in clusters (the Butcher-Oemler effect). (3) For quiescent objects (i.e. ellipticals and S0's), our color indices resolve (in a limited fashion) age and metallicity effects. Since galaxies are a mix of stellar populations of various metallicities, we have adopted a series of multi-metallicity models, calibrated to globular cluster data, which match spectroscopic results. Armed with these indices, our goal is to study the color-magnitude (mass-metallicity) relation in distant clusters and mean age of ellipticals as a function of luminosity (mass).

Figure 1.2 displays the color-magnitude relations (CMR) for A2218 ($z = 0.18$) and A2125 ($z = 0.25$). The CMR is perhaps the most fundamental tool for the analysis of stellar populations in galaxies dating back to Sandage & Visvanathan (1978). The observed relationship in luminous ellipticals has largely been interpreted as reflecting a cooler RGB population (redder color) due to increased metallicity with larger galaxy masses. This has, in large part, been confirmed by numerous absorption line studies (Trager *et al.* 2002) comparing line strengths of features such as $Mg_2$ and Fe with dynamic measures of galaxy mass (i.e. velocity dispersion) and luminosity (a corollary for constant $M/L$). However, there is increasing evidence in the last few years that some of the spread in the CM relation is due to age effects or recent star formation, particularly at the low luminosity end (Poggianti *et al.* 2001).

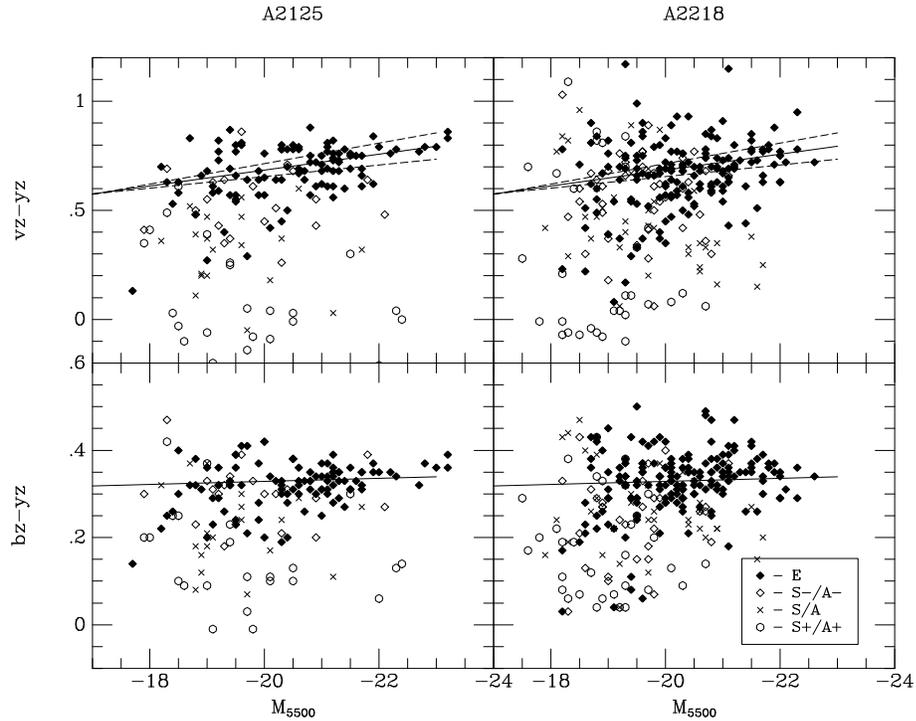

Fig. 1.2. The above two diagrams are the color-magnitude relations for A2218 ($z = 0.18$) and A2125 ($z = 0.25$). The two color indices, $vz - yz$ and $bz - yz$, are shown. The various symbols display the spectrophotometric classifications based on all four filters.

With respect to the Strömgren system, the two color indices, $vz - yz$ and $bz - yz$, versus $M_{5500}$ are shown in Figure 1.2. The various symbols display the spectrophotometric classifications based on all four filters. Points to note: (1) While A2218 is a denser, richer cluster than A2125, both clusters have similar absolute numbers of blue galaxies (for comparison, Coma has no galaxies bluer than $bz - yz = 0.2$ at $M < -19$). Since A2218 is richer, its blue fraction is lower than A2125, although there is clearly a blue population present. (2) The solid lines are fits to Coma and Fornax data (Odell, Schombert and Rakos 2002) and are adequate fits to A2218 and A2125 with some minor, but important differences (see figure on average colors). (3) While $vz - yz$ is more sensitive to metallicity effects than $bz - yz$, $bz - yz$ will still respond to changes in the mean RGB temperature. The lack of any correlation in $bz - yz$ indicates an age effect, such that lower luminosity ellipticals are older (have higher mean stellar ages) then giant cluster ellipticals.

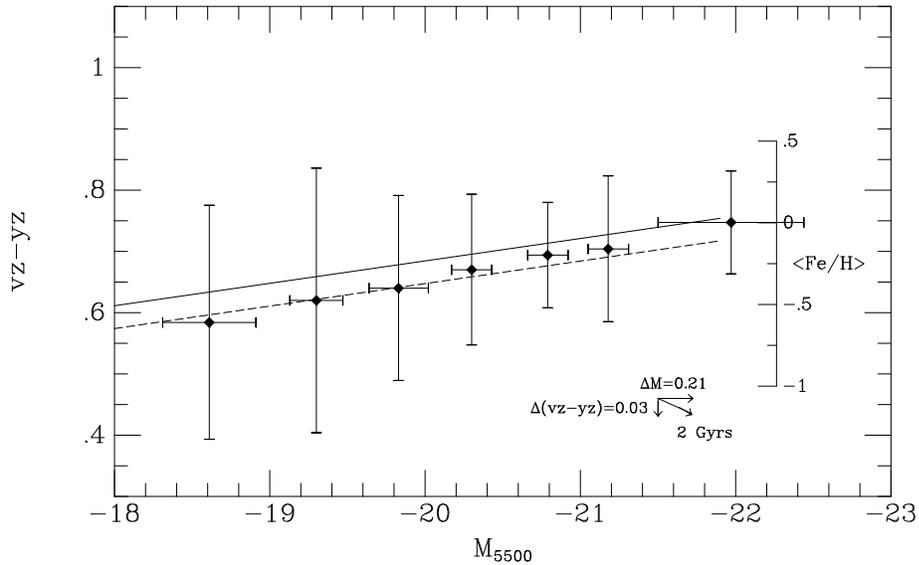

Fig. 1.3. The above diagram displays the average colors of ellipticals (class E only) in luminosity bins of 0.5 mags from -18 to -22.5. The solid line is the best fit Coma/Fornax CMR. The dotted line is the expected change in the CMR for 2 Gyrs of evolution (based on Bruzual & Charlot models) which corresponds to the look-back time for A2125 and A2218 based on Benchmark cosmology values.

While a redshift of 0.2 is a limited lookback time with respect to color evolution, an advantage can be obtained by examining the mean change in the CMR which will minimize the spread in average color due to metallicity. Figure 1.3 displays the average colors of ellipticals in both A2125 and A2218 in bins of 0.5 mags from −18 to −22.5. The colors of both cluster's elliptical populations agree well with passive evolution expectations based on the Bruzual & Charlot models ($\Delta M = 0.21$ and $\Delta(vz - yz) = 0.03$ for 2 Gyrs). Also shown is the mean metallicity scale for the CMR based on multi-metallicity models (see Rakos *et al.* 2001), where evolution effects change the metallicity scale by 0.2 dex by $z = 0.2$.

The age indicator is the $\Delta(bz - yz)$ index. In our Fornax study, $\Delta(bz - yz)$ was determine with respect to a fiducial relationship as given by globular clusters with similar ages. Due to the much higher metallicities found for cluster ellipticals, a comparison to metal-poor globular clusters is problematic as they occupy such different regions of the two-color diagram. As an alternative method, we have calculated the residual $bz - yz$ color based on the distance in the two-color diagram from the 13 Gyr models of Bruzual & Charlot (2002). These models are an excellent fit to the red edge of the color-magnitude relation (see Figure 1.2), and so provide a convenient initial guess at the age of elliptical populations. Since the $\Delta(bz - yz)$ index is a relative age scheme, the choice of this fiducial zero point will not affect our results. Figure 1.4 plots the $\Delta(bz - yz)$ value as a function of luminosity for the ellipticals in Coma and Fornax. The solid line is a linear least squares fit to the Coma/Fornax ellipticals which yields a slope of 0.0038±0.0018 with a correlation coefficient of 0.22. The probability of obtaining this or higher R by chance is 5%.

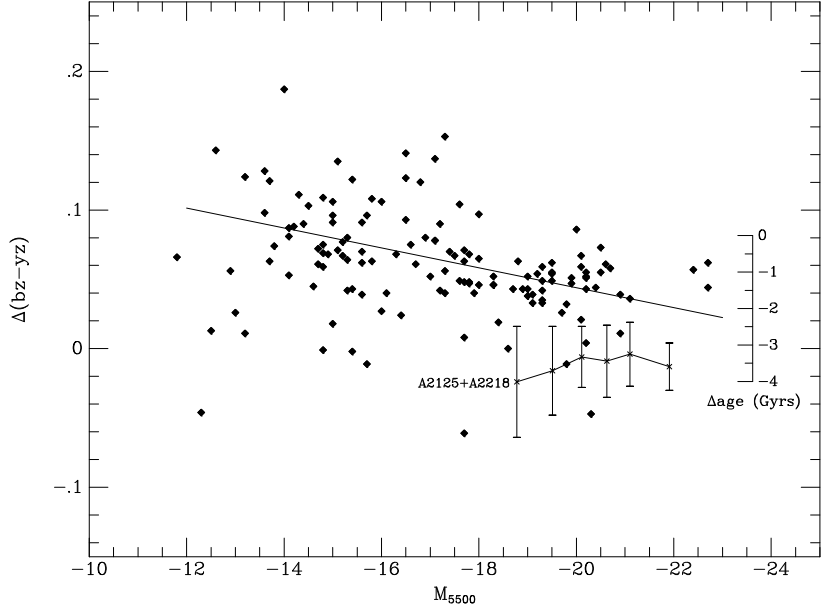

Fig. 1.4. $\Delta(bz-yz)$ index measures the mean age of the underlying stellar population in quiescent objects such as ellipticals. Data for the Coma and Fornax cluster are shown with a clear signature of younger mean age with increasing luminosity. Binned data for A2125 and A2218 are also shown. There is no clear trend with luminosity, although the mean age of ellipticals is 2 Gyrs younger than similar luminosity galaxies in Coma, in agreement with a lookback time of 2 Gyrs for redshift 0.2 clusters.

As can be seen in Figure 1.4, there is a clear trend for redder $\Delta(bz-yz)$ values with lower luminosity for ellipticals. This implies the surprising result that low luminosity ellipticals are older than their higher mass counterparts. As calibrated to globular clusters, a change in 0.25 in $\Delta(bz-yz)$ corresponds to a change of 2 Gyrs in mean stellar age. Thus, by a magnitude of −16, the mean age of a Coma elliptical is 2 Gyrs older than the brighter ellipticals, assumed to be 13 Gyrs by their match to the Bruzual & Charlot 13 Gyr models. This trend of older age with fainter magnitude also merges nicely into the distribution defined by Fornax dwarf ellipticals which have the reddest $\Delta(bz-yz)$ residuals. Interestingly, the three brightest ellipticals in Coma (NGC 4839, 4874, 4889), which lie at cluster subgroup cores, have younger ages than expected, perhaps due to recent cannibalism of star-forming companions.

The averaged data for A2125 and A2218 is also shown in Figure 1.4, binned into six groups of 25 members. The bright end of the distribution agrees well with the expected difference of 2 Gyrs from lookback time. The fainter galaxies appear to decrease in $\Delta(bz-yz)$ (i.e. younger age), however the errors are too large to assign a high statistical weight to this trend. More depth in limiting magnitude to the sample would greatly enhance this analysis, one of our primary goals in the next observing season.

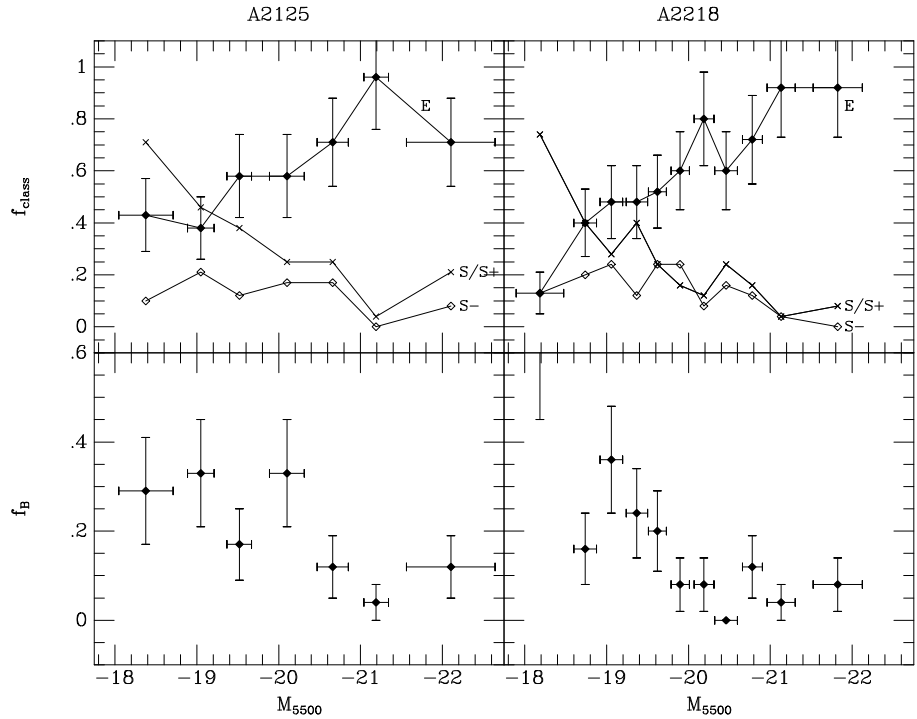

Fig. 1.5. Population fractions for A2125 and A2218, as given by spectrophotometric classification (top panel), and the blue fraction (bottom panel) as a function of absolute luminosity. A2125 displays the classic Butcher-Oemler effect by having a bright blue population of star-forming and starburst galaxies. Both clusters have increasing blue fractions with decreasing luminosity (the dwarf starburst population), however, some of these galaxies are dwarf ellipticals with low metallicity colors.

Lastly, we examine the distribution of the spectrophotometric classes by mass (luminosity). As demonstrated in Rakos *et al.* (2000), the mean star formation rate is defined by spectrophotometric criteria which, in turn, is assigned a spectrophotometric classification. Using a large sample of nearby galaxies, through a principal moment analysis system, allows us to divided our samples into E (old), S- (passive, S0-type), S (disk star formation rates) and S+ (starburst, high star formation rates). While these classes will certainly strong correlate with Hubble type, it is important to remember that this is solely a color criteria and not based on physical morphology. In addition to spectrophotometric classification, we can also assign a fraction of blue galaxies similar to original Butcher-Oemler criteria, although in the rest frame of the cluster without k-corrections.

The distribution of blue fraction, $f_B$, and spectrophotometric class as a function of absolute luminosity is shown in Figure 1.5. With respect to $f_B$, A2125 displays the classic Butcher-Oemler effect by having a bright blue population of star-forming and starburst galaxies. Both clusters have increasing blue fractions with decreasing luminosity (the dwarf starburst population), however, some of these galaxies are likely to be dwarf ellipticals with low metallicity colors masquerading as star-forming galaxies. With respect to spectropho-tometric classes, both A2125 and A2218 have growing starburst populations with lower luminosity. Transitional objects (S-) remain fairly constant with luminosity, which has the implication that the mean age of lower luminosity object decreases due to recent star formation even while the mean age of non-star-forming systems (dwarf ellipticals) increases with decreasing luminosity.

## 1.2    Summary

While spectroscopic studies of distant clusters can provide detailed line indices and kinematic information, narrow band photometry still plays a crucial role in understanding the star formation history, particularly at the faint end of the cluster luminosity function. In our most current study using the Strömgren filter system, we have compared the color properties of two intermediate redshift clusters, A2125 and A2218. While identical in lookback time, A2218 is much more 'evolved' in terms of its global dynamic state and mature red population. A2125 stands as a classic example of a blue Butcher-Oemler cluster. Even though selected to have vastly difference cluster populations, in fact, the trends of age and metallicity are similar for A2125 and A2218. Their color-magnitude (metallicity) and $\Delta(bz-yz)$ (age) relationships follow the same trends as Coma and Fornax when corrected for passive evolution due to lookback time. Their population subclasses, as given by spectrophotometric classification, have identical trends with the sole difference being the number of blue galaxies at the bright end of the luminosity function. This reinforces the idea that the Butcher-Oemler effect primarily involves the most massive galaxies and further study of cluster populations at higher redshifts will be a powerful tool in probing the star formation history of faint cluster galaxies, a test of hierarchical merger scenarios.